# Automatic recognition and detection of aphasic natural speech

*Mara Barberis*[1], *Pieter De Clercq*[1], *Bastiaan Tamm*[2,3], *Hugo Van hamme*[2], *Maaike Vandermosten*[1]

[1] Experimental Otorhinolaryngology (ExpORL), Department of Neurosciences, KU Leuven
[2] Processing Speech and Images (PSI), Department of Electrical Engineering (ESAT), KU Leuven
[3] Laboratory for Cognitive Neurology (LCN), Department of Neurosciences, KU Leuven

`mara.barberis@kuleuven.be`

## Abstract

Aphasia is a language disorder affecting one third of stroke patients. Current aphasia assessment does not consider natural speech due to the time consuming nature of manual transcriptions and a lack of knowledge on how to analyze such data. Here, we evaluate the potential of automatic speech recognition (ASR) to transcribe Dutch aphasic speech and the ability of natural speech features to detect aphasia. A picture-description task was administered and automatically transcribed in 62 persons with aphasia and 57 controls. Acoustic and linguistic features were semi-automatically extracted and provided as input to a support vector machine (SVM) classifier. Our ASR model obtained a WER of 24.5%, outperforming earlier ASR models for aphasia. The SVM shows high accuracy (86.6%) at the individual level, with fluency features as most dominant to detect aphasia. ASR and semi-automatic feature extraction can thus facilitate natural speech analysis in a time efficient manner in clinical practice.

## 1. Introduction

Every year, more than 13.7 million people experience a stroke [1]. As a result, approximately one third of stroke survivors are confronted with aphasia [2]. Aphasia is an acquired language disorder caused by a focal brain lesion in the language-dominant hemisphere, which can affect the person's communicative and social functioning, as well as their quality of life [3]. Current aphasia assessment substantially relies on isolated tasks using single phonemes, words or sentences [4]. However, such artificial tasks (e.g., picture naming, word repetition) lack ecological validity, since natural speech processing requires interaction and higher-level context integration of these isolated stimuli [5]. Clinicians and researchers have therefore advocated to include natural speech in aphasia assessment [6]. Natural speech analysis however requires the time consuming transcribing and subsequent analyzing of a language sample, which can require up to 60 minutes for every minute of speech sampled [7]. Due to this high workload, natural speech analyses have only been limitedly applied in aphasia. In this regard, recent advances in automatic speech recognition and assessment might enable more time efficient natural speech analysis for aphasia. Furthermore, the psychometric properties of natural speech features in aphasia have only been investigated to a limited extent (e.g., [8], [9], implying that there is very little evidence to motivate the choice of one feature over another [9] and insufficient information to use natural speech as a stand-alone assessment tool [10]. The aim of this study is therefore twofold. First, to evaluate ASR performance for aphasic speech, and second, to investigate which natural speech features are relevant to detect aphasia.

## 2. Related work

Currently, several challenges in automated recognition and detection of aphasic natural speech remain. First, there is a relevant decrease in the accuracy of ASR for aphasic speech [11], [12], [13]. This decrease in performance can be caused by aphasia-specific error patterns as well as co-occurring motor speech disorders, which are difficult to capture with conventional ASR [14], [15]. ASR performance furthermore varies widely within persons with aphasia [13], with mild aphasia resulting in better ASR performance compared to more severe aphasia [14], [15]. In addition, ASR systems typically target speech with a known referent, which makes the system suboptimal for natural (connected) speech [15]. Previous studies that applied ASR to aphasia often used tasks with isolated words or sentences and reported 75.7% [16] to 99% [17] agreement between human and ASR transcription. However, when applying ASR to natural aphasic speech, the performance decreases to 37.37% word error rate (WER) [15] or 37.12% [18] to 48.08% [13] syllable error rate (SER). Lastly, ASR advancement is often hindered in low-resource languages like Belgian Dutch. Previous studies applying ASR for aphasia often relied on AphasiaBank [19]. Since AphasiaBank does not contain Dutch data, this study uses a newly collected dataset to evaluate the performance of an ASR model for Dutch aphasic natural speech.

In terms of aphasia detection, 536 natural speech features have been identified [20]. While previous work in the automatic detection of stroke-induced aphasia suggested that mainly linguistic features are relevant to distinguish between aphasia and controls [13], a recent study in persons with primary progressive aphasia found that combined linguistic and acoustic features from natural speech are relevant to distinguish aphasia subtypes [21]. Studies detecting (progressive) aphasia (subtypes) based on linguistic features obtained accuracies ranging from 73% to 100% [22], [23], [24], [25]. Whereas prior research (e.g., [15], [18], [26]) mainly focused on evaluating the robustness of automatically extracted features or predicting aphasia severity from these features, we evaluate the relative importance of each of these features for aphasia detection.

## 3. Methods

### 3.1. Participants and task

The sample consists of 62 Dutch-speaking persons with stroke-induced aphasia (35 male, mean age 69 years) in the chronic phase (≥ 6 months post onset) and 57 neurologically healthy controls (22 male, mean age 65 years). All persons with aphasia (PWA) scored below the cut-off on aphasia severity (ScreeLing [27]) or naming (NBT [28]) tests. Participants had to complete the picture description task from the Comprehensive Aphasia Test [29], which consists of describing a scene in not more than five minutes. Audio was recorded in a quiet space at the participant's home using a BOYA BY-M1 Universal Lavalier Microphone with a sample rate of 16 kHz and a bit-depth of 16. The mean audio duration was 111.04 s (sd = 68.26 s) for persons with aphasia and 65.88 s (sd = 31.28 s) for controls. All participants consented to participate in the study.

### 3.2. Automatic speech recognition

Transcripts were obtained using a Dutch ASR model. This model was trained on 270 hours of Belgian Dutch (the accent matching the test set) speech from the Spoken Dutch Corpus [30] and finetuned on nine hours of natural speech from eighteen cognitively healthy older persons (9 male, mean age 73 years). The model uses a 12-layer conformer encoder and 6-layer transformer [31] decoder. The training is identical to the baseline from Poncelet & Van hamme [32], whereas the encoder is upgraded from a transformer to a conformer module. The performance of this model was evaluated by means of the word error rate (WER) for a subsample of 45 persons with aphasia (25 male, mean age 69 years) and 29 controls (10 male, mean age 63 years). Human annotators manually corrected the ASR generated transcripts and simultaneously labeled unintelligible speech, non-speech sounds and aphasia-specific errors. The latter includes those errors that frequently occur in aphasic speech, being semantic errors, phonemic errors, neologisms, grammatical errors and disfluencies (word repetition, word interruption, (filled) pauses). To ensure conformity of these manual corrections and subsequent labeling, all annotators were trained in using an orthographic transcription protocol [33], adapted for aphasic speech. To evaluate a potential time gain when using ASR for natural speech analysis, all annotators logged the time needed to manually correct and label the speech samples. Since the human annotators used the ASR hypothesis as a reference during annotation, we expect the WER to be lower compared to transcription done from scratch. The WER of an independent ASR model (whisper-large-v3 [34]) is therefore also reported for comparison.

### 3.3. Feature extraction

Natural speech features were extracted in both an automatic and manual manner. The first part of the automatic feature extraction pipeline involved chunking of the audio recording. rVAD [35] was used to find segments of voice activity. Cuts were made in the center of silent segments, such that each chunk duration was as close to 15 seconds as possible (average of 7.4 chunks per aphasic speaker and 4.4 chunks per control speaker). Next, transcripts were obtained for each chunk using the ASR model described in the previous section. Next, Montreal Forced Aligner (MFA) [36] was used to calculate word-level and phone-level timings from the generated transcripts. Finally, we gathered the chunk-wise results to obtain a list of absolute start/end times for each word and phoneme. From this output, automatic features such as word/phone count, number of fillers, and number of short/long pauses were derived. The manually extracted features included counting the occurrences of each of the aphasic errors.

The resulting nineteen linguistic and acoustic features (Table 1) span different subdomains of language. On the lexical level, number of words, Type-Token Ratio (TTR), Brunet's Index (BI) and Honoré's Statistic (HS) were calculated. Since TTR, the number of different words (types) divided by the number of words (tokens), can be biased by text length, BI (a text-length independent measure) and HS (taking into account words only occurring once) were included as well [37]. This is especially relevant when investigating aphasic speech, given that production can strongly vary depending on aphasia severity and that fluency deficits can result in repetitions of a word. On the semantic level, we manually evaluated whether all key concepts depicted in the picture description were produced, and whether semantic errors were present. Phonemic errors, neologisms, grammatical errors and the ability to express causal relations using conjunction words were manually assessed as well. Fluency was included as speech rate (both word and phone rate), long (≥ 400 ms [15], [38]) and short (< 400 ms) pauses, filled pauses, interrupted words, repeated words and repetitions within a word. Finally, unintelligible words were taken into account. All features were normalized for sample length.

Table 1: *Overview of the extracted features*

| Subdomain | Feature | Extraction method |
|---|---|---|
| Lexical | Number of words | Automatic |
|  | Type-Token Ratio (TTR) | Automatic |
|  | Brunet's Index (BI) | Automatic |
|  | Honoré's Statistic (HS) | Automatic |
| Semantic | Key concepts | Manual |
|  | Semantic error | Manual |
| Phonemic | Phonemic error | Manual |
|  | Neologism | Manual |
| Grammatical | Grammatical errors | Manual |
|  | Causal relations | Manual |
| Fluency | Words per minute | Automatic |
|  | Phones per minute | Automatic |
|  | Long pauses | Automatic |
|  | Short pauses | Automatic |
|  | Filled pauses | Automatic |
|  | Word interrupted | Manual |
|  | Word repetition | Manual |
|  | Repetition within a word | Manual |
| Intelligibility | Unintelligible words | Manual |

### 3.4. Aphasia detection

The nineteen natural speech features from Table 1 served as input to a nonlinear support vector machine (SVM) classifier trained to detect aphasia at the individual level. The analyses were performed using the sklearn package (version 1.2.2) implemented in Python (version 3.11). A nonlinear radial basis function kernel SVM with a nested cross-validation approach was used. In the inner cross-validation, the C-hyperparameter and pruning were optimized (accuracy-based) and tested in a validation set using 5-fold cross-validation. Predictions were made on the test set in the outer loop using leave-one-subject-out cross-validation. Evaluation metrics including the receiver operating characteristic (ROC) curve, the area under the curve (AUC), accuracy, F1-score, sensitivity and specificity were computed for the classifier. Model interpretation was performed using SHAP (Shapley Additive exPlanations) analysis [39]. SHAP values quantify the impact of individual features on the SVM's classification. Positive SHAP values indicate features that drive predictions towards the positive class (aphasia), while negative values suggest features favoring the negative class (controls). SHAP values around 0 reflect a minimal impact of that feature on the classification.

## 4. Results

### 4.1. ASR performance

The ASR model had a WER of 24.5% (ranging from 1.3% to 54.9%) for persons with aphasia and 12.0% (ranging from 0% to 34.1%) for controls. For comparison, whisper-large-v3 has a WER of 38.4% (ranging from 5.1% to 102%) for persons with aphasia and 25.0% (ranging from 7.2% to 86.0%) for controls. The largest WER values are due chunk deletions ("no speech" wrongly emitted) and large hallucinations. The WER was significantly increased in aphasia compared to neurologically healthy older controls (U = 1025.0, p < 0.001) and higher WERs were significantly associated with more severe aphasia ($\rho$ = -0.48, p = 0.001). This semi-automated approach for transcription and (both manual and automatic) feature extraction from the picture description task took on average 16.6 minutes (sd = 9.7 minutes) per subject with aphasia.

### 4.2. Aphasia detection

The classifier detected persons with aphasia with an accuracy of 86.6% (Figure 1). Misclassified participants scored significantly higher on the ScreeLing (U = 100.0, p < 0.001) and NBT (U = 96.0, p = 0.002), indicating that misclassified cases were mild cases of aphasia. The area under the curve (AUC) was 90.9%. The SVM had a sensitivity of 79.0% and specificity of 94.7% for aphasia. In Figure 2, natural speech features are ranked according to their impact on the classification of aphasia versus controls (SHAP analysis). Speech rate was ranked highest, meaning this is the strongest discriminative feature, followed by long pauses, causal relations (a proxy for grammatical complexity), repetitions within a word and semantic errors.

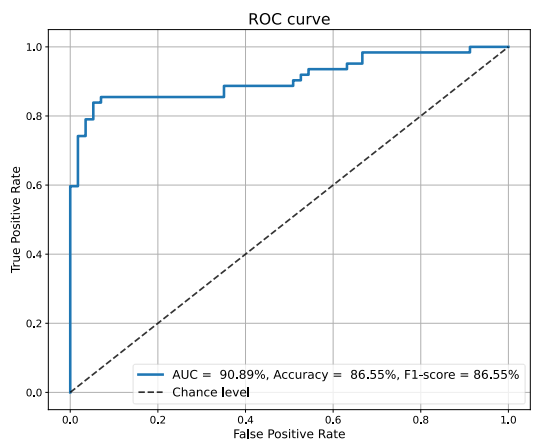

Figure 1: *Support Vector Machine performance*.

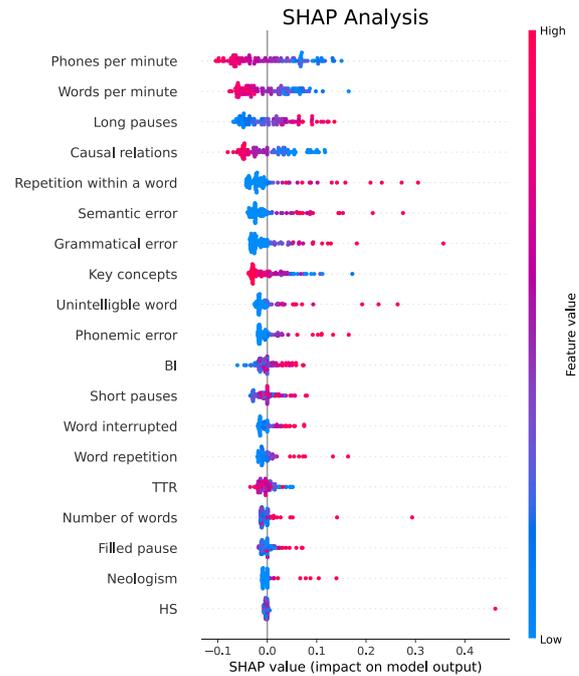

Figure 2: *Feature importance ranking*.

## 5. Discussion

### 5.1. ASR performance for aphasia

Although the WER of our ASR model applied to aphasia (24.5%) is slightly higher than the word error rate of 10%-20% often found in non-pathological natural speech datasets [40], this is still a promising result since disfluencies inherent to pathological speech make it challenging to obtain low WERs. Furthermore, our ASR model outperforms earlier models applied to aphasic natural speech [15] and whisper-large-v3 [34]. However, the wide range in ASR performance (1.3% to 54.9% WER) with increased WER for severe aphasia leaves room for improvement of the model. Implementing ASR in a population of older persons poses additional challenges, since speech and language abilities decrease with (healthy) ageing [41], [42] and older persons more often speak dialect, which can decrease ASR performance [43]. Moreover, post-stroke language deficits (aphasia) often co-occur with post-stroke speech deficits (dysarthria, apraxia of speech). Our sample did not consist of 'pure aphasia' cases, implying that high WERs may also be due to combined speech and language deficits.

### 5.2. Aphasia detection using natural speech features

The support vector machine was able to detect persons with aphasia with a high accuracy of 86.6%. This result was obtained using nineteen natural speech features extracted from an ecologically valid task that poses little burden on the patient. Fluency measures were the dominating features that drove classification of aphasia versus controls. Long pauses seemed to contribute more to aphasia detection compared to short pauses. These long pauses could be an indication of word finding difficulties, a prominent feature of aphasia. Furthermore, both grammatical complexity (causal relations) and accuracy (grammatical errors) were important features to detect aphasia. Within the lexical domain, a text-length independent measure, Brunet's Index, was most suitable for aphasia detection. As expected based on their dominance in aphasia diagnostics, semantic features were included in the top 10 feature ranking. With a sensitivity of 79.0% for aphasia, there is still potential for further improvement of the SVM classifier when including additional linguistic features, such as mean length of utterance, or noun-verb ratio. Leveraging techniques from the field of Natural Language Processing could further automate these linguistic analyses, which were mainly performed manually in the current study.

In future work, the aphasia sample could be divided into stroke survivors with solely language deficits (aphasia) and stroke survivors with both language and speech deficits (aphasia + dysarthria or apraxia of speech). This allows to evaluate whether the acoustic features currently ranked high in the SHAP analysis (phones per minute, words per minutes, pauses) withstand for the purely aphasic group, or whether this effect is mainly driven by the presence of comorbid speech disorders. In addition, it is hypothesized that natural speech features would be more sensitive to detect subtle forms of aphasia compared to current aphasia assessment [44]. For example, in latent aphasia, individuals do not score below the cut-off on aphasia assessment, but they do however report communication difficulties in everyday life. It would be interesting to investigate whether these latent aphasia cases are classified as aphasia or control and which features drive this decision.

### 5.3. Clinical implications

In current aphasia studies, often only transcription or analysis has been automated. The combination of both automatic recognition and detection of aphasic speech however has the potential to significantly lower the workload, making natural speech analysis feasible in clinical practice. Furthermore, acoustic analysis is currently not implemented in standardized aphasia assessment. Nevertheless, our study showed that for example speech rate and long pauses are important features in distinguishing between persons with aphasia and neurologically healthy older persons. This suggests an added value of combining acoustic and linguistic features in the assessment of stroke-induced aphasia, which was also found for progressive aphasia [21].

The cost for persons with aphasia is mostly associated with the time of the clinician [45] and the combined expertise [46] required for a thorough analysis of speech samples. Therefore, automated natural speech analysis is not only an ecologically valid and time efficient, but also a more cost effective manner to assess aphasia. With an average transcription and analysis time of 16.6 minutes for 1-2 minutes of natural aphasic speech, this is four times faster compared to the earlier reported transcription and analysis time of 60 minutes per minute of aphasic speech sampled [7].

## 6. Conclusions

Our findings show that an ASR model finetuned on natural speech is able to transcribe aphasic speech with a WER of 24.5%, which is lower compared to earlier studies. Furthermore, natural speech features derived from an ecologically valid task (picture description) allow for highly accurate aphasia detection. Relevant features to distinguish persons with aphasia from neurologically healthy older persons are speech rate, long pauses, grammatical complexity, repetitions within a word and semantic errors. We conclude that this (semi)automated approach is a promising tool to enable natural speech analysis in a time efficient manner in clinical practice.

## 7. Acknowledgements

We would like to thank all participants and speech and language pathologists involved in this study. This research was financially supported by the Research Foundation Flanders (grant MB: 1SH1Q24N, grant PDC: 1S40122N, grant HVH: S004923N) and the KU Leuven Research Fund (grant C24M/22/025).

## 8. References

[1] P. M. Lindsay *et al.*, "Global Stroke Fact Sheet 2019," *World Stroke Organization (WSO)*, 2019, [Online]. Available: https://www.world-stroke.org/assets/downloads/WSO_Fact-sheet_15.01.2020.pdf

[2] M. Manning, A. MacFarlane, A. Hickey, and S. Franklin, "Perspectives of people with aphasia post-stroke towards personal recovery and living successfully: A systematic review and thematic synthesis," *PLoS One*, vol. 14, no. 3, pp. 1–23, 2019, doi: 10.1371/journal.pone.0214200.

[3] I. Papathanasiou and P. Coppens, *Aphasia and related neurogenic communication disorders*, Second edition. Burlington, MA: Jones & Bartlett Learning, 2017.

[4] M. C. Leaman and L. A. Edmonds, "Assessing Language in Unstructured Conversation in People With Aphasia: Methods, Psychometric Integrity, Normative Data, and Comparison to a Structured Narrative Task," 2021, doi: 10.23641/asha.

[5] L. S. Hamilton and A. G. Huth, "The revolution will not be controlled: natural stimuli in speech neuroscience," *Lang Cogn Neurosci*, vol. 35, no. 5, pp. 573–582, Jun. 2020, doi: 10.1080/23273798.2018.1499946.

[6] J. H. Azios, B. Archer, and J. B. Lee, "Detecting behavioural change in conversation: procedures and preliminary data," *Aphasiology*, vol. 35, no. 7, pp. 961–983, 2021, doi: 10.1080/02687038.2020.1812031.

[7] L. Armstrong, M. Brady, C. Mackenzie, and J. Norrie, "Transcription-less analysis of aphasic discourse: A clinician's dream or a possibility?," *Aphasiology*, vol. 21, no. 3–4, pp. 355–374, Mar. 2007, doi: 10.1080/02687030600911310.

[8] M. Boyle, "Stability of word-retrieval errors with the Aphasiabank stimuli," *Am J Speech Lang Pathol*, vol. 24, no. 4, pp. S953–S960, Nov. 2015, doi: 10.1044/2015_AJSLP-14-0152.

[9] M. Pritchard, K. Hilari, N. Cocks, and L. Dipper, "Psychometric properties of discourse measures in aphasia: acceptability, reliability, and validity," *Int J Lang Commun Disord*, vol. 53, no. 6, pp. 1078–1093, Nov. 2018, doi: 10.1111/1460-6984.12420.

[10] M. Pritchard, K. Hilari, N. Cocks, and L. Dipper, "Reviewing the quality of discourse information measures in aphasia," *Int J Lang Commun Disord*, vol. 52, no. 6, pp. 689–732, 2017, doi: 10.1111/1460-6984.12318.

[11] B. MacWhinney and D. Fromm, "Language Sample Analysis With TalkBank: An Update and Review," *Front Commun (Lausanne)*, vol. 7, Apr. 2022, doi: 10.3389/fcomm.2022.865498.

[12] Y. Qin, T. Lee, S. Feng, and A. P. Hin Kong, "Automatic speech assessment for people with aphasia using TDNN-BLSTM with multi-task learning," in *Proceedings of the Annual Conference of the International Speech Communication Association, INTERSPEECH*, International

[13] Y. Qin, T. Lee, and A. P. Hin Kong, "Automatic Speech Assessment for Aphasic Patients Based on Syllable-Level Embedding and Supra-Segmental Duration Features," in *ICASSP, IEEE International Conference on Acoustics, Speech and Signal Processing - Proceedings*, Institute of Electrical and Electronics Engineers Inc., Sep. 2018, pp. 5994–5998. doi: 10.1109/ICASSP.2018.8461289.

Speech Communication Association, 2018, pp. 3418–3422. doi: 10.21437/Interspeech.2018-1630.

[14] D. Le and E. M. Provost, "Improving automatic recognition of aphasic speech with AphasiaBank," in *Proceedings of the Annual Conference of the International Speech Communication Association, INTERSPEECH*, International Speech and Communication Association, 2016, pp. 2681–2685. doi: 10.21437/Interspeech.2016-213.

[15] D. Le, K. Licata, and E. Mower Provost, "Automatic quantitative analysis of spontaneous aphasic speech," *Speech Commun*, vol. 100, pp. 1–12, Jun. 2018, doi: 10.1016/j.specom.2018.04.001.

[16] K. J. Ballard, N. M. Etter, S. Shen, P. Monroe, and C. T. Tan, "Feasibility of Automatic Speech Recognition for Providing Feedback During Tablet-Based Treatment for Apraxia of Speech Plus Aphasia," 2019, doi: 10.23641/asha.

[17] A. Jacks, K. L. Haley, G. Bishop, and T. G. Harmon, "Automated Speech Recognition in Adult Stroke Survivors: Comparing Human and Computer Transcriptions," *Folia Phoniatrica et Logopaedica*, vol. 71, no. 5–6, pp. 286–296, Oct. 2019, doi: 10.1159/000499156.

[18] Y. Qin, T. Lee, and A. P. H. Kong, "Automatic Assessment of Speech Impairment in Cantonese-Speaking People with Aphasia," *IEEE Journal on Selected Topics in Signal Processing*, vol. 14, no. 2, pp. 331–345, Feb. 2020, doi: 10.1109/JSTSP.2019.2956371.

[19] B. MacWhinney, D. Fromm, M. Forbes, and A. Holland, "Aphasiabank: Methods for studying discourse," *Aphasiology*, vol. 25, no. 11, pp. 1286–1307, 2011, doi: 10.1080/02687038.2011.589893.

[20] L. Bryant, A. Ferguson, and E. Spencer, "Linguistic analysis of discourse in aphasia: A review of the literature," *Clin Linguist Phon*, vol. 30, no. 7, pp. 489–518, 2016, doi: 10.3109/02699206.2016.1145740.

[21] C. Themistocleous, B. Ficek, K. Webster, D. B. den Ouden, A. E. Hillis, and K. Tsapkini, "Automatic subtyping of individuals with primary progressive aphasia," *Journal of Alzheimer's Disease*, vol. 79, no. 3, pp. 1185–1194, 2021, doi: 10.3233/JAD-201101.

[22] K. C. Fraser *et al.*, "Automated classification of primary progressive aphasia subtypes from narrative speech transcripts," *Cortex*, vol. 55, no. 1, pp. 43–60, 2014, doi: 10.1016/j.cortex.2012.12.006.

[23] S. Lukic *et al.*, "Discriminating nonfluent/agrammatic and logopenic PPA variants with automatically extracted morphosyntactic measures from connected speech," *Cortex*, vol. 173, pp. 34–48, Apr. 2024, doi: 10.1016/j.cortex.2023.12.013.

[24] L. Wagner, M. Zusag, and T. Bloder, "Careful Whisper -- leveraging advances in automatic speech recognition for robust and interpretable aphasia subtype classification," Aug. 2023.

[25] G. Chatzoudis, M. Plitsis, S. Stamouli, A.-L. Dimou, A. Katsamanis, and V. Katsouros, "Zero-Shot Cross-lingual Aphasia Detection using Automatic Speech Recognition," Apr. 2022, [Online]. Available: http://arxiv.org/abs/2204.00448

[26] M. Day, R. K. Dey, M. Baucum, E. J. Paek, H. Park, and A. Khojandi, "Predicting Severity in People with Aphasia: A Natural Language Processing and Machine Learning Approach," in *Proceedings of the Annual International Conference of the IEEE Engineering in Medicine and Biology Society, EMBS*, Institute of Electrical and Electronics Engineers Inc., 2021, pp. 2299–2302. doi: 10.1109/EMBC46164.2021.9630694.

[27] E. G. Visch-Brink, M. van de Sandt-Koenderman, and H. El Hachioui, *ScreeLing*. Bohn Stafleu Van Loghum, 2010.

[28] L. Van Ewijk, L. Dijkhuis, M. Hofs-van Kats, M. Hendrickx, M. Wijngaarden, and C. De Hilster, *Nederlandse benoemtest (NBT)*. Bohn Stafleu van Loghum, 2018.

[29] E. Visch-Brink *et al.*, *Comprehensive Aphasia Test - Nederlandstalige bewerking (CAT-NL)*. Pearson, 2015.

[30] N. Oostdijk, "The Spoken Dutch Corpus: Overview and first evaluation," *Proceedings of LREC-2000, Athens*, vol. 2, Jan. 2000.

[31] A. Vaswani *et al.*, "Attention Is All You Need," Jun. 2017.

[32] J. Poncelet and H. Van hamme, "Learning to jointly transcribe and subtitle for end-to-end spontaneous speech recognition," in *Proceedings SLT 2022*, IEEE, 2022.

[33] Taalunie, "Protocol voor Orthografische Transcriptie." 2000.

[34] A. Radford, J. W. Kim, T. Xu, G. Brockman, C. McLeavey, and I. Sutskever, "Robust Speech Recognition via Large-Scale Weak Supervision," Dec. 2022.

[35] Z.-H. Tan, A. kr. Sarkar, and N. Dehak, "rVAD: An unsupervised segment-based robust voice activity detection method," *Comput Speech Lang*, vol. 59, pp. 1–21, Jan. 2020, doi: 10.1016/j.csl.2019.06.005.

[36] M. McAuliffe, M. Socolof, S. Mihuc, M. Wagner, and M. Sonderegger, "Montreal Forced Aligner: Trainable Text-Speech Alignment Using Kaldi," in *Interspeech 2017*, ISCA: ISCA, Aug. 2017, pp. 498–502. doi: 10.21437/Interspeech.2017-1386.

[37] A. Ntracha, D. Iakovakis, S. Hadjidimitriou, V. S. Charisis, M. Tsolaki, and L. J. Hadjileontiadis, "Detection of Mild Cognitive Impairment Through Natural Language and Touchscreen Typing Processing," *Front Digit Health*, vol. 2, 2020, doi: 10.3389/fdgth.2020.567158.

[38] S. V. S. Pakhomov *et al.*, "Computerized analysis of speech and language to identify psycholinguistic correlates of frontotemporal lobar degeneration," *Cognitive and Behavioral Neurology*, vol. 23, no. 3, pp. 165–177, Sep. 2010, doi: 10.1097/WNN.0b013e3181c5dde3.

[39] S. M. Lundberg and S.-I. Lee, "A Unified Approach to Interpreting Model Predictions," 2017. [Online]. Available: https://github.com/slundberg/shap

[40] P. Szymański *et al.*, "WER we are and WER we think we are," Oct. 2020, [Online]. Available: http://arxiv.org/abs/2010.03432

[41] M. Kuruvilla-Dugdale, M. Dietrich, J. D. McKinley, and C. Deroche, "An exploratory model of speech intelligibility for healthy aging based on phonatory and articulatory measures," *J Commun Disord*, vol. 87, Sep. 2020, doi: 10.1016/j.jcomdis.2020.105995.

[42] P. Tremblay, J. Poulin, V. Martel-Sauvageau, and C. Denis, "Age-related deficits in speech production: From phonological planning to motor implementation," *Exp Gerontol*, vol. 126, Oct. 2019, doi: 10.1016/j.exger.2019.110695.

[43] S. Feng, B. M. Halpern, O. Kudina, and O. Scharenborg, "Towards inclusive automatic speech recognition," *Comput Speech Lang*, vol. 84, p. 101567, Mar. 2024, doi: 10.1016/j.csl.2023.101567.

[44] E. Gleichgerrcht *et al.*, "Neural bases of elements of syntax during speech production in patients with aphasia," *Brain Lang*, vol. 222, Nov. 2021, doi: 10.1016/j.bandl.2021.105025.

[45] N. Boyer, N. Jordan, and L. R. Cherney, "Implementation Cost Analysis of an Intensive Comprehensive Aphasia Program," *Arch Phys Med Rehabil*, vol. 103, no. 7, pp. S215–S221, Jul. 2022, doi: 10.1016/j.apmr.2020.09.398.

[46] M. Jacobs and C. Ellis, "Understanding the Economics of Aphasia: Recent Findings from Speech and Language Research," *Seminars in Speech and Language*, vol. 43, no. 3. Thieme Medical Publishers, Inc., pp. 198–207, Jun. 01, 2022. doi: 10.1055/s-0042-1749132.